\documentclass[article]{IEEEtran}
\usepackage[utf8]{inputenc}
\usepackage{amsmath}
\usepackage{graphicx}
\usepackage{circuitikz}
\usepackage{cite}
\usepackage{xcolor}

\usepackage[greek,english]{babel}
\newcommand{\gpi}{\textrm{\greektext p}}
\renewcommand{\pi}{\gpi}
\newcommand{\ju}{\mathrm{j}}
\newcommand{\eu}{\mathrm{e}}
\newcommand{\hf}{\mathrm{h}^{(2)}}
\newcommand{\ka}{\rho}
\newcommand{\za}{Z_\mathrm{a}}
\newcommand{\zta}{\tilde{Z}_\mathrm{a}}

\title{Distortion Analysis for the Assessment of LTI and non-LTI Transmitters}
\author{Kurt Schab \IEEEmembership{Member, IEEE}, Amritpal Singh, and Nicole Bohannon
\thanks{Manuscript received  \today; revised \today.}
\thanks{K.~Schab and A.~Singh are with the Department of Electrical Engineering, Santa Clara University, Santa Clara, CA, USA (e-mail: kschab@scu.edu).}
\thanks{N. Bohannon is with the Laboratory for Physical Sciences, USA.}}
\begin{document}

\maketitle

\begin{abstract}
    Bounds on the frequency domain behavior of electrically small antennas are adapted to assess the time domain distortion or fidelity achievable by simple linear time-invariant (LTI) systems.   Rigorous expressions for a TM$_{01}$ spherical shell are used as a direct analog to the well known Chu limit on an antenna's bandwidth-efficiency product. These expressions are shown to agree with results obtained using narrowband single resonance approximations, allowing for the analysis of arbitrary small dipole-like structures through bounds on their single frequency Q-factor and efficiency.  The resulting expressions are used as a basis for assessing the performance gains of electrically small non-LTI (e.g., direct antenna modulation) transmitters for which Q-factor and impedance bandwidth are not defined but which can be analyzed directly in the time domain via distortion.
\end{abstract}


\IEEEpeerreviewmaketitle

\section{Introduction}
\label{sec:intro}
\IEEEPARstart{B}{ounds} on the performance of electrically small antennas provide meaningful estimates to the limiting trends in key metrics for wireless systems.  Bounds on Q-factor\footnote{Readers are directed to \cite{Schab2017energy} for a detailed discussion of Q-factor and its various definitions for radiating systems.} (closely related to impedance bandwidth), and radiation efficiency are two of the most studied parameters, with mature methodologies available for their computation on a variety of objects.  From analytic methods on spheres \cite{Chu1948,Pfeiffer_FundamentalEfficiencyLimtisForESA} to advanced numerical procedures for arbitrary geometries \cite{capek2017minimization,gustafsson2019trade}, the general trends are that both Q-factor and efficiency are dramatically constrained as the size of an electrically small antenna is reduced.

One possibility to circumvent these bounds is through the use of non-linear or time-varying (non-LTI) antennas, which are not necessarily constrained in the same ways as their linear time-invariant (LTI) counterparts.  Several promising approaches have been reported using switched matching networks \cite{Galejs1963,Vallese1972,Gamble1973,Xu2006,Zhu2014,manteghi2017,schab2019pulse,manteghi2019}, switched DC supplies \cite{Merenda2006,Manteghi2016}, or time-varying tuning and matching elements on an antenna itself \cite{Keller2010,Salehi2013} to successfully transmit broadband signals from electrically small antennas .  However, lacking classically defined impedances or Q-factors, these time-varying systems cannot be directly compared to analogous LTI systems by the common practice of comparing impedance bandwidth and continuous wave (CW) efficiency.  This presents a major obstacle in the fair, replicable evaluation of non-LTI devices and their relative performance tradeoffs compared to conventional systems. 

The goal of this work is to reformulate well-known behavioral trends of LTI electrically small dipole antennas into forms compatible with time domain metrics applicable to non-LTI systems.  These metrics are based heavily on those developed for assessing signal distortion (or equivalently, fidelity) in ultra-wideband (UWB) transmit systems \cite{farr1992,Lamensdorf1994} which deal directly with the signals radiated by an arbitrary source.  By moving to this methodology, we remove all dependence of assumptions on the LTI / non-LTI nature of a transmitter.

We begin by summarizing distortion, as used in this paper, in Section~\ref{sec:dist-and-eff}.  This is followed in Section~\ref{sec:mappings} by development of field-circuit mappings necessary for the study of distortion due to both spherical TM$_{01}$ radiators (exact, specific) and single resonance electrically small dipole type antennas (approximate, general).  Several numerical examples studying the accuracy of the approximate single resonance model and comparison of various simulated electrically small antennas are presented in Section~\ref{sec:examples}.  Finally we apply the developed methodology to assess the performance of measured conventional (LTI) and direct antenna modulation (DAM, non-LTI) systems reported in \cite{schab2019pulse}.



\section{Distortion}
\label{sec:dist-and-eff}
The narrow impedance bandwidth of an LTI electrically small antenna distorts a transmitted or received signal, reducing the efficacy of information transfer.  Typically, this distortion is not considered directly, but rather by proxy through the antenna's impedance bandwidth.  In an effort to study electrically small antennas in ways applicable both to LTI (conventional) and non-LTI (time-varying systems for which impedance bandwidth is not defined), we revert to the study of signals produced by a radiating antenna directly in the time domain. 

Let the distortion measured between an ideal signal $x_\mathrm{i}$ and a realized signal $x$ be defined as \cite{Lamensdorf1994}
\begin{equation}
\label{eq:d-td}
d(x_\mathrm{i},x) = \min_\tau \int_{-\infty}^\infty \left| \hat{x}_\mathrm{i}(t) - \hat{x}(t-\tau) \right|^2\mathrm{d}t.
\end{equation}
where $\hat{\cdot}$ denotes the normalization
\begin{equation}
	\hat{f}(t) =\frac{f}{\left[\int_{-\infty}^{\infty} |f|^2\mathrm{d}t\right]^{1/2}}.
\end{equation}
Note the optimization over time delay $\tau$, with the true distortion defined by the minimum possible value.  Parseval's Theorem allows this to be alternatively written in the frequency domain as 
\begin{equation}
\label{eq:d-fd}
d(X_\mathrm{i},X) = \min_\tau \int_{-\infty}^\infty \left| \hat{X}_\mathrm{i}(\omega) - \hat{X}(\omega)\mathrm{e}^{-\mathrm{j}\omega\tau} \right|^2\mathrm{d}\omega.
\end{equation}
From distortion, fidelity $F$ may also be calculated, as the two metrics are closely related by
\begin{equation}
\label{eq:fid}
    F= 1 -\frac{d}{2}
\end{equation}
where the minimization over time delay $\tau$ is implicitly taken on whichever quantity is directly computed \cite{Lamensdorf1994}.  By these conventions, a linear system with a non-zero transfer function produces zero distortion (unity fidelity) under CW excitation. 

If there exist multiple degrees of freedom in each observable, for example, two polarizations of a radiated field, the integrand in \eqref{eq:d-td} and \eqref{eq:d-fd} can be generalized to the norm of the difference between the ideal and realized signals in vector form, i.e.,
\begin{equation}
\label{eq:d-td-gen}
d(\boldsymbol{x}_\mathrm{i},\boldsymbol{x}) = \min_\tau \int_{-\infty}^\infty \left| \boldsymbol{x}_\mathrm{i}(t) - \boldsymbol{x}(t-\tau) \right|^2\mathrm{d}t.
\end{equation}

If measurement distortion (e.g., channel distortion, limited receiver bandwidth) or preconditioning (e.g., polarization selection) is known and representable by the linear operator $\mathcal{L}$, the same formulation for distortion can be applied using \cite{Lamensdorf1994}
\begin{equation}
\label{eq:op-transform}
	x = \mathcal{L}y,\quad x_\mathrm{i} = \mathcal{L}y_\mathrm{i}
\end{equation}
where $y$ and $y_\mathrm{i}$ are the realized and ideal transmit signals, and now $x$ and $x_\mathrm{i}$ are the corresponding observables. 

Though distortion is used extensively in the characterization of UWB transmitters, the impact of the bandwidth and efficiency limitations of small antennas on distortion bounds has not been previously described.  Our goal here is to do so in order to provide a benchmark against which the distortion properties of non-LTI systems can be compared.  We proceed by developing operational relationships between ideal signals (e.g., input voltage to an antenna) and some produced output (e.g., a normalized radiated field) in order to calculate distortion in terms of parameters bounded for LTI systems.  In this way, we project the effect of an LTI system's Q-factor and efficiency onto the distortion metric which is compatible for comparison with non-LTI systems for which Q-factor, impedance bandwidth, and other frequency domain parameters are not defined.  The rest of this paper uses distortion as the metric of interest, though all data and results can be recast in terms of fidelity using \eqref{eq:fid}.

\section{Radiated Fields from Electrically Small TM$_{01}$ Antennas}
\label{sec:mappings}
The goal of this section is to derive expressions for calculating the fields radiated by a small TM$_{01}$ antenna under specific (LTI) matching conditions.  These in turn can be used to calculate distortion as defined in Sec.~\ref{sec:dist-and-eff} under arbitrary excitation.  A spherical radiator with a well-defined equivalent circuit is first studied before applying approximations to generalize the calculations to the study of any structure (antenna or current distribution) with a Q-factor and efficiency defined at a single frequency.

\subsection{Spherical Shell}
In a seminal paper studying the physical bounds of omnidirectional antennas, Chu established an equivalent circuit to model the impedance and radiation behavior of a dipole-like (TM$_{01}$) spherical antenna \cite{Chu1948}.  In terms of the input current $I$ to this equivalent circuit, the radiated electric field $\boldsymbol{E}$ is
\begin{equation}
    \boldsymbol{E} = \hat{\boldsymbol{\theta}}\frac{1}{\ka\hf_1}\sqrt{\frac{3\eta_0}{8\pi}}\sin\theta \frac{\eu^{-\ju kr}}{r} I.
\end{equation}
Here $\ka$ denotes the electrical size $ka = 2\pi a/\lambda$ and $\hf_1$ is the first order spherical Hankel function of the second kind evaluated with argument $\ka$.  Without loss of generality, we consider only broadside ($\theta=\pi/2$) radiated fields, drop the fixed constant $\sqrt{3\eta_0/(8\pi)}$, suppress the far field propagation term, and consider only the scalar magnitude of the $\theta$-polarized field, i.e.,
\begin{equation}
    E = \frac{I}{\ka\hf_1}.
    \label{eq:sphere-ei}
\end{equation}
The equivalent antenna input impedance of the TM$_{01}$ radiator is defined as
\begin{equation}
    Z_\mathrm{a} = \frac{\ju(\ka\hf_1)'}{\ka\hf_1} = \frac{\ka^4 - \ju\ka}{\ka^4+\ka^2}
\label{eq:z1}
\end{equation}
and can be readily implemented using a three element RLC circuit \cite{Chu1948}.  Here $(\cdot)'$ denotes a derivative taken with respect to electrical size $\rho$.  Note that this impedance represents only an outward going wave impedance.  Internal fields can easily be included using parallel networks \cite{thal1978exact} with the effect of increased Q-factors in the calculation of electrically small spheres in Sec.~\ref{sec:examples}.

Suppose the antenna (equivalent circuit) is tuned to resonance at $\ka_0$ by a series inductor $L$.  Further, a series resistance $R_\Omega$ is introduced to represent ohmic losses.  This resistance can be determined based on material parameters and electrical size by various models.  Because of the intricate nature of its definition and scaling (e.g., skin depth model or surface resistance) we leave this resistance as a free parameter to tune at will. With these alterations,  the tuned, lossy impedance model is written as 
\begin{equation}
    \zta = \za + \ju\ka X_0/\ka_0 + R_\Omega,
\end{equation}
where the tuning inductance is scaled in terms of its reactance \mbox{$X_0 = \mathrm{Im}~\tilde{Z}_\mathrm{a}(\ka_0)$} at the tuned frequency.  A non-dispersive source with impedance \mbox{$R_0 = \mathrm{Re}~\tilde{Z}_\mathrm{a}(\ka_0)$} is connected to the antenna circuit.  For a given excitation $V_\mathrm{s}$, the radiated field is then given by
\begin{equation}
    E = \frac{V_\mathrm{s}}{\ka \hf_1(R_0+\zta)}.
    \label{eq:sphere-ev}
\end{equation}
The above expression is general in the sense that no approximations have been made regarding the electrical size of the spherical radiator, only that it radiates solely in the TM$_{10}$ mode.  Using some form of this electric field as an observable quantity, the relation in \eqref{eq:sphere-ev} can be applied to calculate distortion from a sphere of arbitrary size ($\ka$), arbitrary loss parameters ($R_\Omega$), and under arbitrary excitation ($V_\mathrm{s}$); where the ideal observable $x_\mathrm{i}$ may be the excitation itself.


\subsection{Arbitrary Electrically Small TM$_{01}$ Radiators}
Severe Q-factor and efficiency limits arise in the electrically small limit ($\ka\rightarrow 0$).  Here we simplify \eqref{eq:sphere-ev} in this limit and make narrowband, stationary approximations to allow for the calculation of the radiated field $E$ with knowledge only of a radiator's tuned Q-factor and efficiency at a given operating frequency.

The current-field relation in \eqref{eq:sphere-ei} simplifies to the expected time-derivative form in the electrically small limit, i.e.,
\begin{equation}
E = -\ju \ka I.
\end{equation}
Using this simplification to represent the general behavior of all small dipole-like radiators and the definition of the reflection coefficient 
\begin{equation}
    \varGamma = \frac{\zta - R_0}{\zta + R_0},
    \label{eq:gamma}
\end{equation}
the radiated field may be written as 
\begin{equation}
    E = \frac{\ju \ka V_\mathrm{s}}{2R_0}\left(1-\varGamma\right).
    \label{eq:esa-ev}
\end{equation}
Assuming that the resonant system can be approximated locally in frequency as a series RLC circuit \cite{Yaghjian2005}, the reflection coefficient can be rewritten using \eqref{eq:gamma} as a function parameterized by the system's resonant radiation Q-factor $Q$ and efficiency $\eta$
\begin{equation}
    \varGamma = \frac{\ju Q\eta (\ka-\ka_0) / \ka_0}{1+\ju Q\eta (\ka-\ka_0) / \ka_0}.
    \label{eq:gamma}
\end{equation}
Here $Q$ denotes only the radiation Q-factor where as $Q\eta$ is the total system Q-factor including loss.  In this RLC approximation, the ohmic and radiation losses are implicitly assumed to be stationary in frequency.

The expressions in \eqref{eq:esa-ev} and \eqref{eq:gamma} represent a framework for calculating the frequency domain radiated fields for an electrically small dipole-like antenna specified by a radiation Q-factor and efficiency at a single frequency.  Arbitrary excitation $V_\mathrm{s}$ may be applied and loss models may be parameterized by the total system efficiency $\eta$.  Given that closed form expressions are available for the radiation Q-factor and efficiency of a TM$_{01}$ spherical current distribution, the strength of the stationary narrowband approximations can be easily assessed.  Similarly, many techniques allow for the calculation of optimal Q-factor and optimal efficiency currents on arbitrarily shaped objects (e.g., \cite{capek2017minimization,gustafsson2019trade}), meaning that \eqref{eq:esa-ev} and \eqref{eq:gamma}, in conjunction with the metrics defined in Sec.~\ref{sec:dist-and-eff}, can be applied to the study of the optimal distortion and efficiency achievable by arbitrarily shaped electrically small dipole-like radiators.

\subsection{Removal of Time Derivative Term}
The expressions \eqref{eq:sphere-ev} and \eqref{eq:esa-ev} suggest that the source voltage and radiated electric field may be candidates for the ideal and measured signals used in computing distortion via \eqref{eq:d-td}.  However, we note that even in the case of perfect broadband matching, there exists a time derivative term between these two quantities.  In an effort to obtain a comparison that reduces to zero distortion in the ideally matched case, either the source voltage must be multiplied by this derivative term $\mathrm{j}\ka$ or the radiated field must be integrated to remove this term.  We opt for the latter in this study to avoid issues related to the unbounded nature of the differentiation operator.  Thus in applying equation \eqref{eq:d-td} to calculate distortion, we take for the ideal signal $x_\mathrm{i} = \mathcal{F}^{-1}\left[V_\mathrm{s}\right]$ and the measured signal $x= \mathcal{F}^{-1}\left[E/(\mathrm{j}\ka)\right]$ where $E$ is the radiated field calculated either by \eqref{eq:sphere-ev} (exact sphere) or \eqref{eq:esa-ev} (approximate single resonance model).  This is similar to the treatment discussed in \cite{farr1992}.


\subsection{Discussion of Parameter Dependencies}
The rigorous calculation of the electric field produced by a spherical shell \eqref{eq:sphere-ev} depends on many parameters: input signal type, antenna materials, antenna efficiency $\eta$, and electrical size $\ka$.  By contrast, it can be observed that, for signals classified by some nominal bandwidth $B$ (e.g., $99\%$ power occupied bandwidth), all of these parameters can be collapsed into a single product $Q\eta B$ using the single resonance model in \eqref{eq:esa-ev}.  Note that here $B$ represents the bandwidth of the input signal (not the antenna bandwidth). Thus, the complete term $Q\eta B$ is proportional to the quotient of the signal and antenna impedance bandwidths \cite{Yaghjian2005}. 

This implies that all small antennas which fit the assumptions of the single resonance dipole model transmitting a signal of bandwidth $B$ can be classified by a single number $Q\eta B$ regardless of their specific Q-factor, efficiency, or the bandwidth of the signal.  Thus, the distortion for a specific signal type and bit sequence can be plotted as a function of this single composite variable.  The resulting curve contains the distortion performance of any and all single resonance antennas transmitting that particular message.  The shape of this curve depends on the exact message being transmitted for short signals but converges as the message length is increased and the power spectral density of the input signal approaches that of an infinite sequence of the chosen modulation type.

At this point it is necessary to highlight a few of the practical advantages of collapsing several antenna metrics into the combined parameter $Q\eta B$.  First, to predict realized distortion under a given signal bandwidth $B$, there is no need to evaluate an antenna's radiation Q-factor or efficiency, rather only the product $Q\eta$ which can be easily estimated from an analytic, numerical, or measured antenna's input reflection coefficient \cite{Yaghjian2005}.  Second, the effect of broadbanding an antenna via resistive loading can quickly be assessed by calculating the associated shift in $Q\eta B$ and moving along the single resonance model curve.  Finally, if a non-LTI antenna is capable of operating in an LTI mode with no changes to hardware (e.g., with a dynamic switch held static as done in \cite{schab2019pulse}), then its effective value of $Q\eta B$ can be determined.  The realized distortion in both LTI and non-LTI modes of such a transmitter can then be measured directly in the time domain and compared to the single resonance model presented here.  Alternative methods for comparing non-LTI data using these metrics are discussed in Sec.~\ref{sec:dam}.


\section{Example Calculations on LTI Systems}
\label{sec:examples}
In this section, we apply distortion calculations to assess LTI structures.  In anticipating the analysis of datasets collected from non-LTI transmitters in \cite{schab2019pulse}, we calculate distortion using pseudorandom bit sequence (PRBS) on-off-keyed (OOK) excitations where the data rate is defined by the number $N$ of carrier periods per symbol.  The nominal bandwidth of these signals is defined here as $B = 1 / N$.  Additional examples using differentiated Gaussian and modulated Gaussian pulses are also included.  Throughout this section, $ka$ denotes the electrical size at the tuned frequency, i.e., $\ka_0$.

\subsection{Assessment of Stationary, Narrowband Approximations}
\begin{figure}
    \centering
    \includegraphics[width=3.25in]{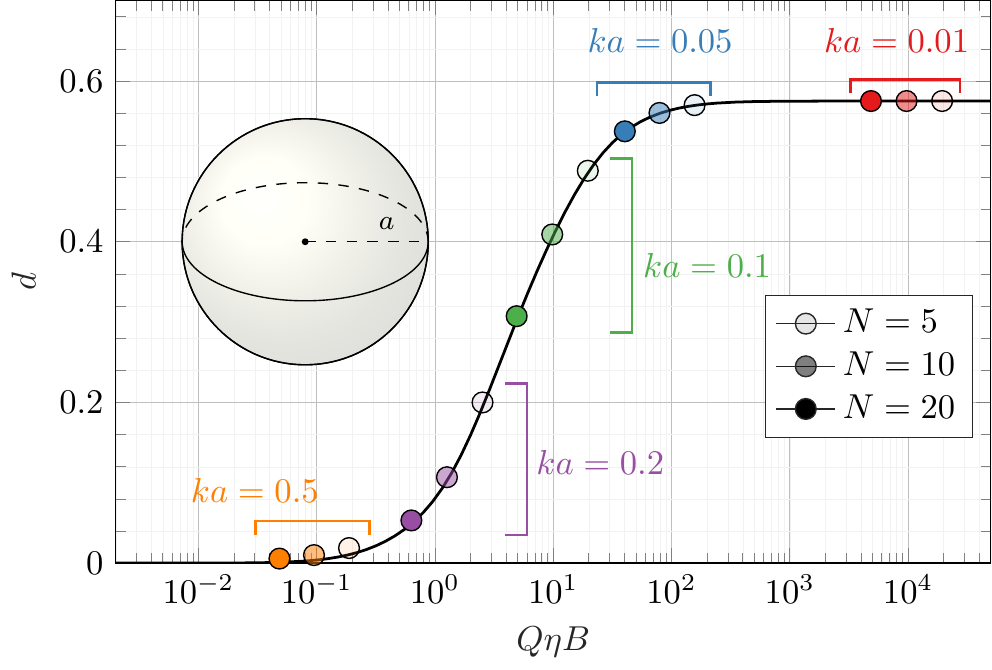}
    \caption{Comparison of rigorous sphere distortion calculations (points) with those based on the single resonance model (solid line).  Bracketed points are generated for a given electrical size $ka$, while opacity denotes the number of cycles $N$ per bit in the input 128-bit OOK PRBS.}
    \label{fig:sphere-comparison}
\end{figure}

Here we compare the distortion resulting from the rigorous expressions derived for a spherical shell with those from the approximate single resonance model.  A 128-bit PRBS sequence is used to excite spherical radiators of varying electrical size, with the radiated field calculated via \eqref{eq:sphere-ev}.  Frequency dependent losses were not considered, instead $R_\Omega$ was selected to enforce a CW radiation efficiency of $\eta = 10\%$ at the tuned carrier frequency for each electrical size.  At each frequency, the tuned factor $Q\eta$ was calculated from the differential impedance bandwidth \cite{Yaghjian2005}.  Data from this calculation are overlaid with a curve of distortion values obtained by sweeping $Q\eta$ in the single resonance model \eqref{eq:esa-ev} in Fig.~\ref{fig:sphere-comparison}.  Excellent agreement is observed for all tested electrical sizes.

\begin{figure}
    \centering
    \includegraphics[width=3.25in]{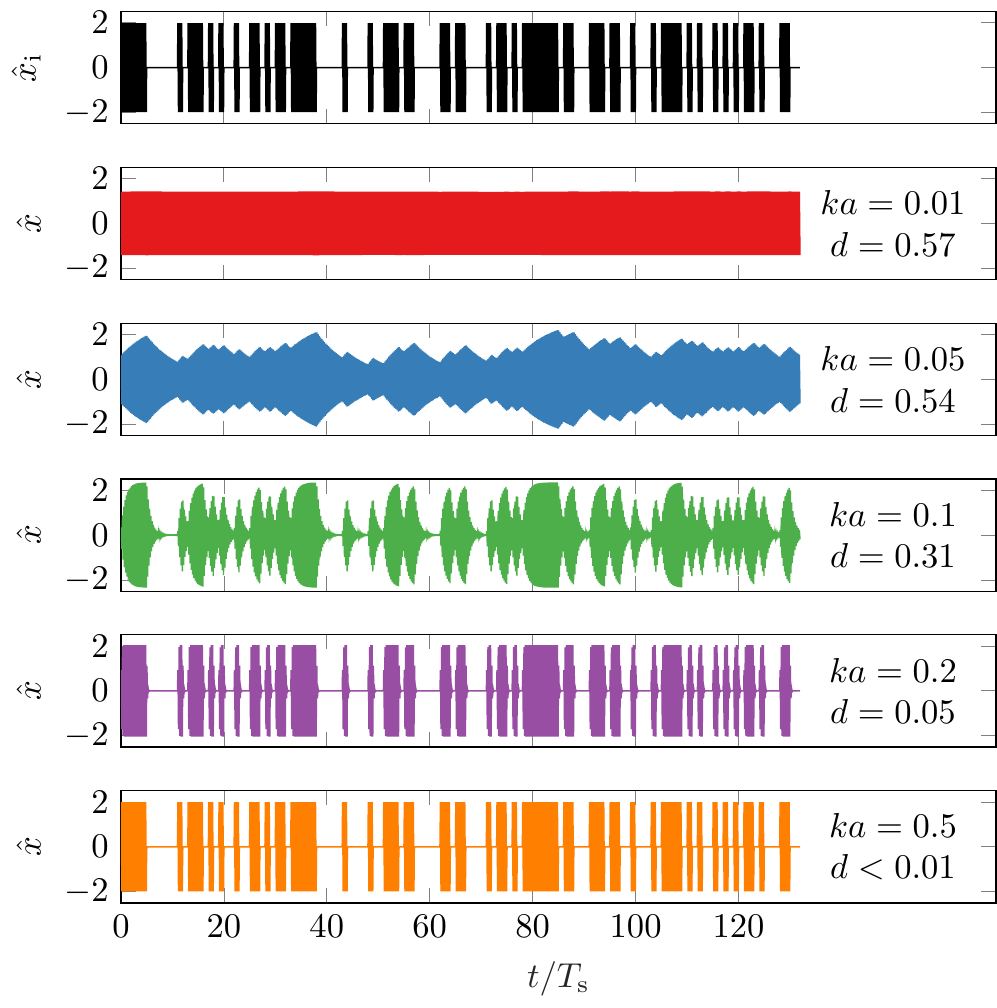}
    \caption{Normalized time-domain signals $\hat{x}$ for points corresponding to \mbox{$N = 20$} in Fig.~\ref{fig:sphere-comparison}.  The ideal signal $\hat{x}_\mathrm{i}$ is shown in the top panel.  Time axis normalized to the symbol length $T_\mathrm{s}$.}
    \label{fig:sphere-td}
\end{figure}

Many time domain signals may yield the same distortion values.  The level of tolerable distortion in a communications link would depend on many parameters related to the receiver architecture.  Making judgements on what is or is not an acceptable level of distortion is beyond the scope of this work, however we do note that link resilience and capacity would nearly always improve with lowered distortion.  For reference, in Fig.~\ref{fig:sphere-td} we plot time domain signals produced by a small selection of the data points in Fig.~\ref{fig:sphere-comparison} along with their corresponding values of distortion.

\begin{figure}
    \centering
    \includegraphics[width=3.25in]{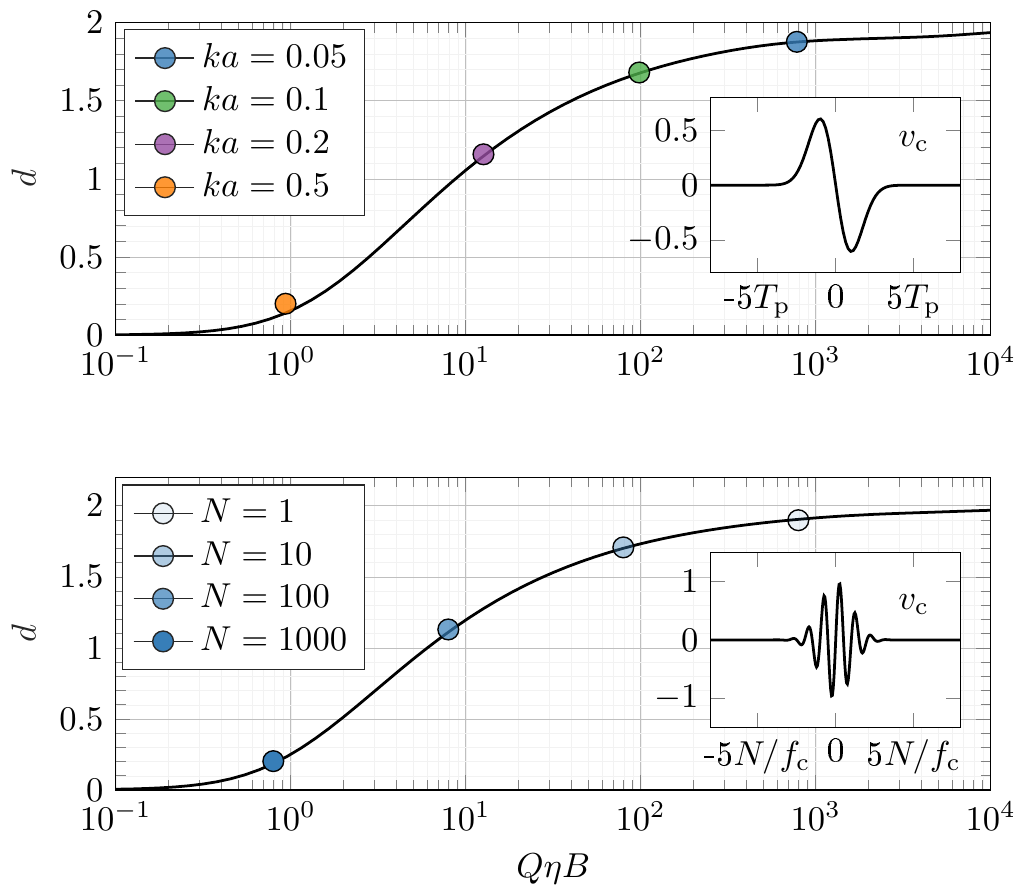}
    \caption{Pulse distortion from a sphere via rigorous calculation (points) and single resonance model (curve).  Differentiated Gaussian pulses for varying sphere sizes (top) and modulated Gaussian pulses for fixed sphere size and varying pulse width (bottom) are shown.}
    \label{fig:sphere-pulses}
\end{figure}

Additionally, we can examine the validity of the single resonance model under extremely broadband conditions using traditional UWB pulses.  Rather than the OOK PRBS signal used throughout the rest of this paper, Fig.~\ref{fig:sphere-pulses} shows the same comparison as Fig.~\ref{fig:sphere-comparison} using differentiated Gaussian 
\begin{equation}
    v_\mathrm{s} = -\frac{t}{T_\mathrm{p}}\mathrm{e}^{-\frac{1}{2}\left(t/T_\mathrm{p}\right)^2}
\end{equation}
and modulated Gaussian 
\begin{equation}
    v_\mathrm{s} = \sin\left(2\pi f_\mathrm{c} t\right)\mathrm{e}^{-\frac{1}{2}\left(tf_\mathrm{c}/N\right)^2}
\end{equation}
pulses \cite{jin2015theory}.  The width of the differentiated Gaussian pulses $T_\mathrm{p}$ corresponds to a peak frequency component located at the tuned frequency of a spherical shell with swept electrical size.  The pulse repetition period is $2048T_\mathrm{p}$.  In the case of the modulated Gaussian pulses, the pulse width is written in terms of the carrier period and is varied by the parameter $N$ and the pulse repetition rate is $2048/f_\mathrm{c}$, where $f_\mathrm{c}$ corresponds to the modulation frequency and the tuned frequency of a spherical shell of size $ka = 0.05$.  In each of these analyses, $\eta = 10\%$.  Excellent agreement is observed.  The nominal signal bandwidth is normalized to $B = 1$ for the differentiated Gaussian and $B = 1/N$ for the modulated Gaussian pulses.

\subsection{Analysis of Optimal Currents on Arbitrary Objects}

With the single resonance model, distortion may be predicted for any system with a Q-factor and efficiency defined at a single frequency.  This includes non-driven current distributions on arbitrary structures, extending the general approach used for spheres to the analysis of any arbitrary geometry.  Currents optimal in radiation Q-factor and efficiency are readily obtained using a method of moments formulation \cite{capek2017minimization}.  In many ways, these optimal, non-driven current distributions are exactly analogous to the spherical harmonics used by Chu, though their behavior is not obtained via analytic expansion, but rather by deterministic convex optimization methods.

\begin{figure}
    \centering
    \includegraphics[width=3.25in]{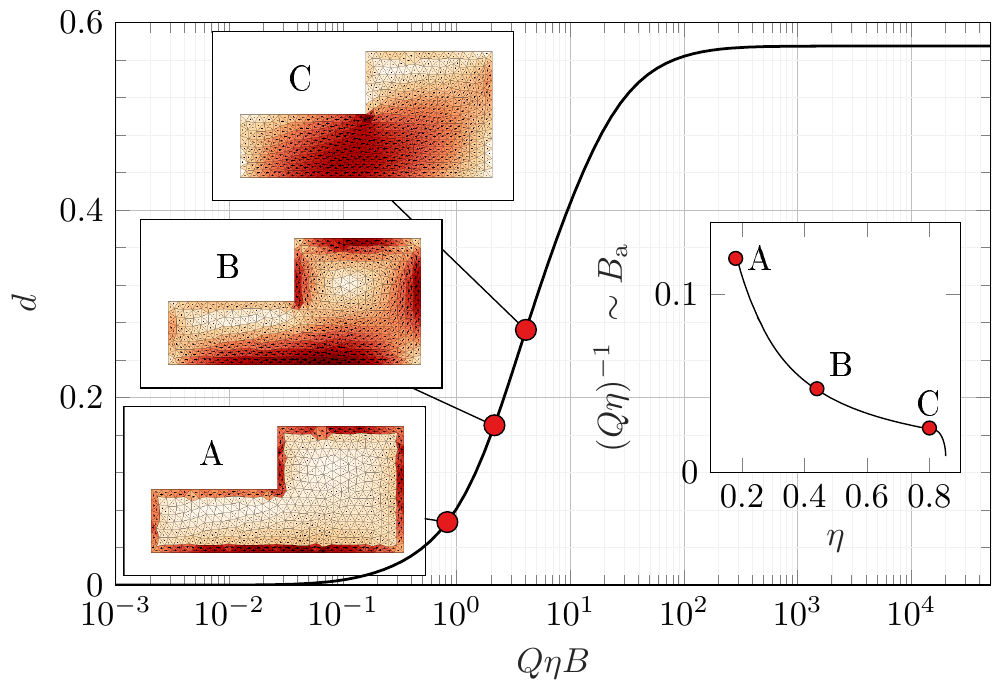}
    \caption{Distortion calculated for Pareto-optimal currents on an L-shaped plate$^\mathrm{1}$.  Three points along the bandwidth-efficiency Pareto front (inset) are selected and the corresponding distortion values are calculated using the single resonance model.}
    \label{fig:ell}
\end{figure}

The tradeoff between optimal radiation Q-factor ($Q$) and optimal efficiency may be calculated via a Pareto analysis~\cite{gustafsson2019trade}.  Here we conduct this analysis for an L-shaped plate\footnote{Aspect ratio $2:1$, notch extends exactly halfway into each dimension.} with electrical size $ka = 0.5$, homogeneous surface resistivity \mbox{$R_\mathrm{s} = 0.1~\Omega$}, and without the explicit constraint of self resonance (i.e., lossless tuning implied for non-resonant solutions).  The results are plotted in the inset of Fig.~\ref{fig:ell} as the tradeoff between $(Q\eta)^{-1}$ (proportional to antenna bandwidth $B_\mathrm{a}$) and efficiency $\eta$.  Three points are selected at various positions along the tradeoff curve and the properties of the associated optimal current distributions are used to calculate distortion of a 128-bit PRBS OOK signal with $N=10$.  The resulting points are shown along side these current distributions in Fig.~\ref{fig:ell}.  Note that, unlike the preceding analysis of a sphere, these data points are computed directly from the single resonance model and lie exactly on the single resonance model curve, drawn in black.

This analysis demonstrates how distortion analysis may be combined with the study of optimal currents to predict \textit{a priori} the ideal tradeoff between message distortion and radiator efficiency.  This is particularly relevant for the analysis of non-LTI systems where it is prudent to assess how improvements in signal fidelity compare versus the gains that could have been achieved by resistively broadbanding a comparable LTI transmitter.



\subsection{Analysis of Driven Antennas}

In the preceding examples, non-driven optimal current distributions were assessed using hypothetical equivalent circuit models.  By contrast, here we demonstrate the calculation of distortion from several driven antennas evaluated by full wave simulations and compare the obtained data to curves predicted by the single resonance model.

\begin{figure}
    \centering  \includegraphics[width=3.25in]{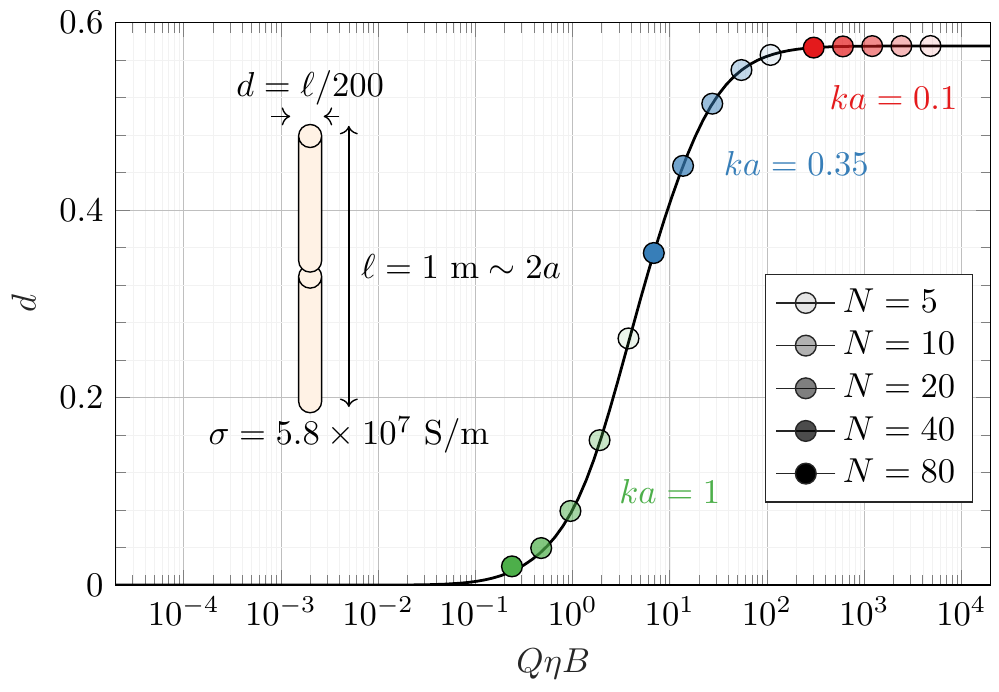}
    \caption{Comparison of full wave distortion calculations on a copper dipole simulated using the method of moments (points) with those based on the single resonance model (solid line).}
    \label{fig:dipoles}
\end{figure}

We begin with a thin wire dipole, depicted in Fig.~\ref{fig:dipoles}.  The dipole is simulated using NEC2++ \cite{nec}, a method of moments solver, to obtain broadband input impedance and radiation data.  These data are then used to numerically tune the dipole to a carrier frequency represented by an electrical size $ka$.  In each case considered here, the dipole is tuned below its natural resonance using a series inductance.  The tuned system is driven by a matched source using the same class of OOK signals as in the previous examples with bandwidths classified by the number of cycles per symbol $N$.  For each signal considered, the broadside co-polarized field is computed, integrated, and compared to the driving voltage via calculation of distortion.  Additionally, at each tuned frequency, the parameter $Q\eta$ is estimated from simulated impedance data using the model in \cite{Yaghjian2005}.  This factor, combined with the nominal signal bandwidth yields the coordinate $Q\eta B$ with which the simulated distortion can be compared to the single resonance distortion model. Fig.~\ref{fig:dipoles} shows excellent agreement between the fullwave dipole simulation and single resonance distortion model over three electrical sizes and five data rates, with each parameter spanning at least one order of magnitude.

\begin{figure}
    \centering
    \includegraphics[width=3.25in]{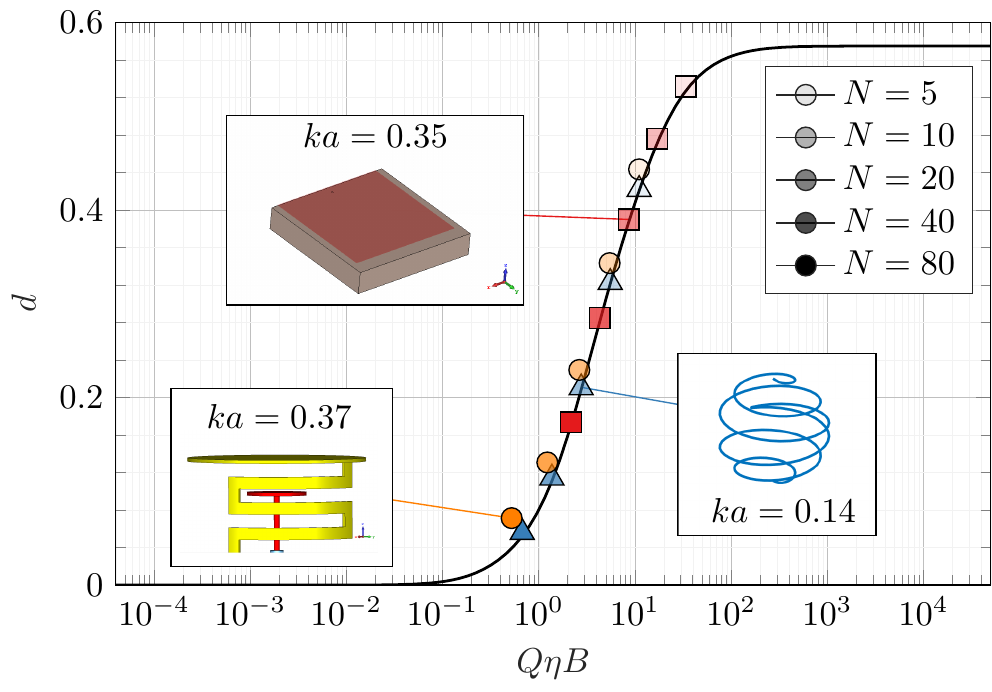}
    \caption{Comparison of full wave distortion calculations using three self-resonant antennas$^\mathrm{2}$ (points) with those based on the single resonance model (solid line).}
    \label{fig:sr-antennas}
\end{figure}

Several self resonant structures\footnote{Dimensions and materials are listed in Appendix A.} are subjected to the same simulation and distortion calculation procedure, with results shown in Fig.~\ref{fig:sr-antennas}.  A spherical helix, a quarterwave patch on a finite dielectric substrate, and a capacitively coupled circular patch \cite{vanniekerk2013analysis} are examined.  The latter two antennas were simulated over an infinite perfect electric conductor (PEC) ground plane using CST Microwave Studio.  As in the previous example, the factor $Q\eta$ was again estimated from simulated impedance data, though here each antenna is driven with a nominal $50~\Omega$ source.  All transmitted fields were calculated in the polarization and direction of maximum gain at the self resonance frequency.  Despite the relatively high complexity of these antennas, we observe excellent agreement between the simulated distortion and the predicted single resonance curve.  

\section{Assessment of a non-LTI Transmitter}
\label{sec:dam}
With the accuracy of the single resonance model verified in previous examples, we move to our primary motivation for studying distortion: assessing the relative performance of LTI and non-LTI transmitters.

As stated in Sec.~\ref{sec:intro}, while previous efforts to demonstrate non-LTI transmitters' ability to outperform comparable LTI transmitters have provided promising results, reporting of the realized benefits of non-LTI systems has been largely visual or qualitative.  This is due in no small part to the fact that frequency domain quantities, such as impedance bandwidth, are not directly defined for non-LTI systems and thus cannot be used as performance metrics as is done with LTI systems.  Here we use distortion, which can be measured directly in the time domain for either LTI or non-LTI systems, to compare the performance of the non-LTI OOK transmitter reported in \cite{schab2019pulse}.

The transmitter studied in \cite{schab2019pulse} is based on the switched matching network architecture described in \cite{Galejs1963}.  As such, it is capable of operating in conventional (LTI) and direct antenna modulation (DAM, non-LTI) modes without changes to the physical hardware.  Thus, the transmitter can be characterized by a value of the parameter $Q\eta$ measured when operating in LTI modes.  Note that not all non-LTI systems may be characterized in this way, but as we shall point out later this does not prevent the use of distortion analysis.

\begin{figure}
    \centering
    \includegraphics[width=3.25in]{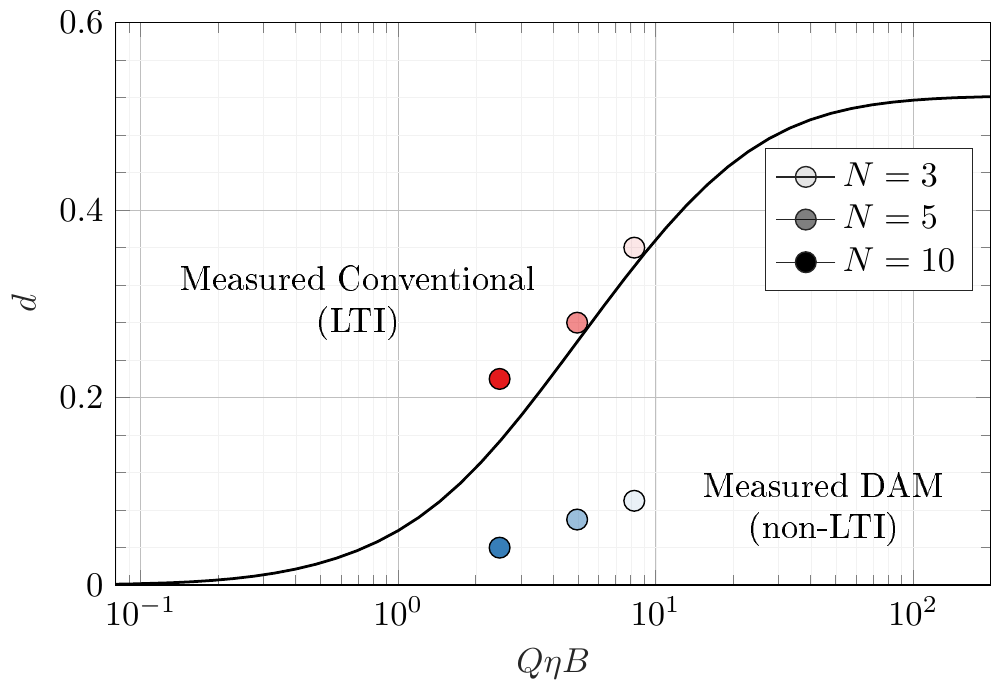}
    \caption{Comparison of OOK LTI distortion bounds with measured non-LTI transmitter performance.  Here $N\sim 1/B$ is a metric of symbol length.  The same 64-bit OOK sequence was used for the analytic calculation and the measured transmissions. }
    \label{fig:dam}
\end{figure}

Raw over the air time domain data from the experiments conducted in \cite{schab2019pulse} were used to calculate distortion for 64-bit PRBS OOK signals in conventional and DAM modes.  Channel inversion was used to isolate distortion due to transmitter mismatch, see \cite{schab2019pulse} for details.  The resulting data are plotted in Fig.~\ref{fig:dam}, where good agreement between the conventional transmitter mode and the single resonance model is observed.  For all three transmissions using the DAM transmitter mode, the distortion is clearly reduced below that of the single resonance model, in line with the visual assessment of corresponding eye diagrams made in \cite{schab2019pulse}.


\begin{figure}
    \centering
    \includegraphics[width=3.25in]{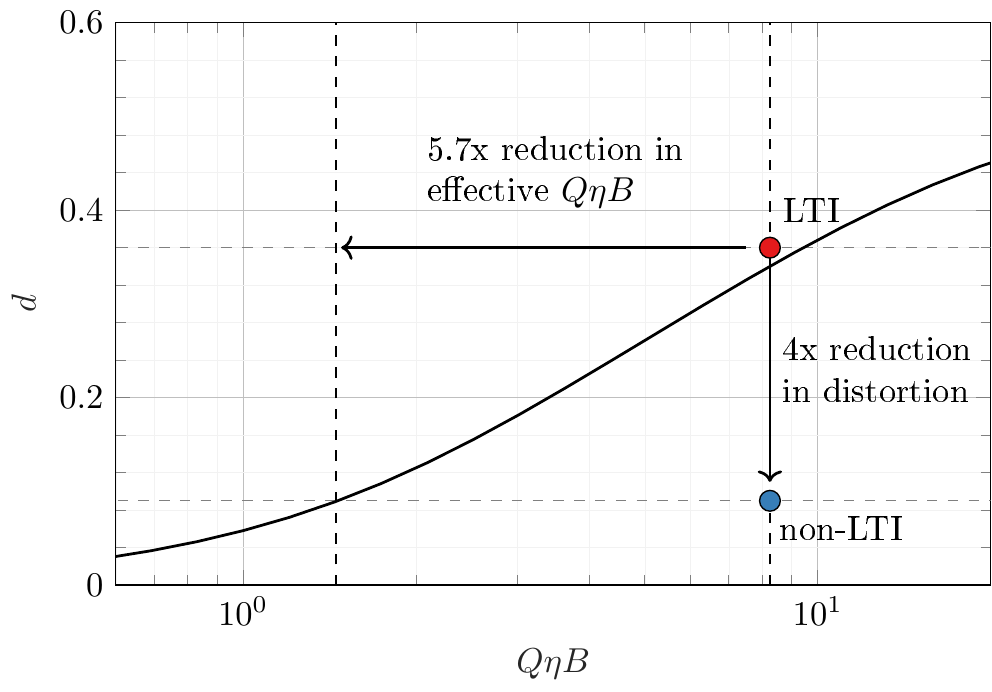}
    \caption{Comparison of OOK LTI distortion bounds with measured non-LTI transmitter performance.  The same 64-bit OOK sequence was used for the analytic calculation and the measured transmissions. }
    \label{fig:dam-detail}
\end{figure}

Beyond qualitative comparison, the use of distortion analysis enables the quantification of gains afforded by this non-LTI transmitter.  Fig.~\ref{fig:dam-detail} shows a detail view of the self resonance curve and measured distortion from the conventional and DAM transmitter modes with $N=3$.  Starting from the indicated LTI measurement, we observe that the non-LTI system provides a four-fold reduction in distortion (from $0.36$ to $0.09$).  To achieve equivalent distortion for the same transmission, the LTI system would require resistive broadbanding amounting to a reduction in efficiency (or gain, assuming constant directivity) by a multiplicative factor of $0.17$ or equivalently $-7.6~\mathrm{dB}$.  Conversely, since the LTI and non-LTI systems have the same efficiency and the signal bandwidths are identical in both measurements, the non-LTI system can be interpreted as having an ``effective Q-factor'' a multiplicative factor of $0.17$ lower than that of the LTI system, or equivalently an ``effective bandwidth'' that is a multiplicative factor of $5.7$ larger.  

Neither gain nor Q-factor is necessarily defined for the non-LTI transmitter. However, this example demonstrates how distortion may be used as an intermediate quantity to determine the non-LTI transmitter's potential benefits in terms of these critical system parameters.  Additionally, while there is no non-trivial lower bound on distortion for non-LTI systems, the curve containing the distortion properties of all single resonance LTI antennas helps illustrate the relative gains and performance parameters afforded by the use of any particular non-LTI transmitter implementation.

\section{Conclusions}

Electric field relations for a spherical shell and single resonance dipole-like antennas are cast in forms compatible with the calculation of distortion.  We have provided several examples justifying the approximation that all LTI, electrically small, dipole-like radiators are well characterized by a single curve of distortion versus the composite Q-factor, efficiency, and signal bandwidth parameter $Q\eta B$.  This finding is used to compare the performance of a non-LTI transmitter by obtaining quantitative measures of its behavior relative to the bandwidth and / or efficiency of a comparable LTI system.  

Going forward, further studies into quantifying the efficiency and directive nature (intentional or unintentional) of non-LTI radiators in order to complete a framework for assessing their integration into communications or sensing systems.  Furthermore, the procedure presented in this paper will allow for the more rigorous analysis of previously proposed theoretical and experimental non-LTI transmitters.

\appendices
\section{Dimensions of Example Self-Resonant Antennas}

\subsection{Spherical Helix}
The five-turn spherical helix is simulated using NEC2++, a method of moments based solver.  The simulated helix is made of a cylindrical wire of radius $r_\mathrm{w} = 0.1~\mathrm{mm}$ with conductivity $\sigma = 5.8\cdot10^{7}~\mathrm{S/m}$.  The radius of the bounding sphere is \mbox{$r = 1~\mathrm{m}$} and the helix is described by the curve $(x(t),~y(t),~z(t))$ with coordinates
\begin{equation}
    z = -r + 2rt,\quad\quad
    \rho = \sqrt{r^2-z^2}
\end{equation}
\begin{equation}
    x = \rho \cos (10\pi t),\quad\quad
    y = \frac{z}{|z|}\rho \sin (10\pi t)
\end{equation}
where $t \in [0,1]$.  The structure is discretized into 175 segments and fed at the segment crossing $t=0.5~(z = 0~\mathrm{m})$.  The helix is resonant near $f = 6.6~\mathrm{MHz}~(ka = 0.14)$.

\subsection{Quarterwave Patch}
The square quarterwave patch is simulated using the finite element method in CST.  The ground plane and patch are made of PEC and the ground plane is of infinite extent. There is a shorting wall the entire extent of the patch on the side closest to the feed point.  Parameters for this antenna are given in Table~\ref{tab:diel}.

\subsection{Dual Reactively Loaded Monopole (DRLM)}
The DRLM is based on the horizontally meandered PIFA from \cite{vanniekerk2013analysis}, where detailed dimensions may be obtained.  The circumscribing sphere for this antenna (and its image below the infinite ground) has a radius of $22~\mathrm{mm}$.  Simulations of this antenna are carried out in CST with an infinite PEC ground plane.  The self resonant frequency is near $810~\mathrm{MHz}~(ka = 0.37)$. 

\begin{table}[]
    \centering
    \begin{tabular}{l|c}
    \textit{Parameter} & \textit{Value}\\ \hline
        Substrate dimensions & $9$~mm \\
        Patch dimensions & $8$~mm \\
        Shorting wall width & $8$~mm\\
        Feed distance from shorting wall & $3.4$~mm \\
        Substrate relative permittivity & $9.8$ \\
        Feed type & SMA coaxial \\ 
        Self resonant frequency & $2.978~\mathrm{GHz}$\\
    \end{tabular}
    \caption{Quarterwave patch parameters.}
    \label{tab:diel}
\end{table}

\bibliographystyle{IEEEtran}
\bibliography{main}

\end{document}